\documentclass[conference]{IEEEtran}
\IEEEoverridecommandlockouts
\usepackage{cite}
\ifCLASSINFOpdf
  \usepackage[pdftex]{graphicx}
  \DeclareGraphicsExtensions{.pdf,.jpeg,.png}
\else
  \usepackage[dvips]{graphicx}
  \DeclareGraphicsExtensions{.eps}
\fi
\usepackage[cmex10]{amsmath}
\usepackage{amssymb}  
\usepackage{steinmetz}
\graphicspath{{./figures/}}
\usepackage{amsthm}

\usepackage{algorithm,algorithmic}
\usepackage{cancel}

\begin{document}

\title{Rapid and Accurate Changepoint Detection of Power System Forced Oscillations \\
\thanks{This material is based upon work supported by the National Science Foundation under Grant No. 1944689}
}

\author{\IEEEauthorblockN{Luke Dosiek}
\IEEEauthorblockA{
Department of Electrical, Computer \\
and Biomedical Engineering\\
Union College\\
Schenectady, NY 12309\\
Email: dosiekl@union.edu}
\and
\IEEEauthorblockN{Akaash Karn}
\IEEEauthorblockA{
Niskayuna High School\\
Niskayuna, NY 12309\\
Email: karnakaash@gmail.com}
\and 
\IEEEauthorblockN{Frank Liu}
\IEEEauthorblockA{
Bronx High School of Science\\
Bronx, NY 10468\\
Email: frankl77@nycstudents.net}
}

\maketitle

\begin{abstract}

This paper describes a new approach for using changepoint detection (CPD) to estimate the starting and stopping times of a forced oscillation (FO) in measured power system data. As with a previous application of CPD to this problem, the pruned exact linear time (PELT) algorithm is used. However, instead of allowing PELT to automatically tune its penalty parameter, a method of manually providing it is presented that dramatically reduces computation time without sacrificing accuracy. Additionally, the new algorithm requires fewer input parameters and provides a formal, data-driven approach to setting the minimum FO segment length to consider as troublesome for an electromechanical mode meter. A low-order ARMAX representation of the minniWECC model is used to test the approach, where a 98\% reduction in computation time is enjoyed with high estimation accuracy. 

\end{abstract}

\section{Introduction}
\label{sec:Intro}
It is a well-known phenomenon that when an electromechanical mode mode meter analyzes power grid measurements that contain a forced oscillation (FO), care must be taken to avoid biasing the results toward 0\% damping \cite{vanfretti2012effects,myers2013effects,trudnowski2016shape,JimFO,urmillaFO,Pmaps,LukeGM19,LukeFhatLetter}. Several works have demonstrated that by incorporating the FO into the model structure being estimated by the mode meter, one can avoid this issue \cite{JimFO,urmillaFO,GM23}. Common to these approaches is that they require highly accurate estimates of the FO parameters.

Recent work in \cite{GM23} described methods of estimating FO amplitude, frequency, and phase that were asymptotically unbiased and with minimum variance. To achieve this, results from \cite{GM22} were used which applied changepoint detection (CPD) to the task of estimating the samples where the FO started and stopped. While accurate, the overall estimation task was throttled by the CPD algorithm, which relied on an iterative process that was slow to converge. Also problematic was that the CPD algorithm had an upper limit on the number of CPs that could be estimated, and it required the tuning of a threshold parameter to classify CPs as FO starts or stops.

This paper describes an improved method of using CPD to estimate the FO start and stop samples that 1) reduces the computational speed by 98\% without sacrificing accuracy, and 2) removes both the upper limit on the number of CPS and the classification threshold.

\section{Background}
\label{sec:Background}
\subsection{The Small-Signal Model}
Under ambient conditions, a power grid measurement such as frequency, power, etc., is filtered and downsampled to produce $N$ samples of a signal that is well-modeled as an autoregressive moving average (ARMA) process $x$ 
\begin{equation}
	x[k] = \frac{C(q)}{A(q)}e[k] 
	\label{eqn:ARMAdefn} 
\end{equation}
where $e$ is zero-mean Gaussian White Noise (GWN), $k$ is the sample index, $q$ is the delay operator such that $q^{-n}x[k]=x[k-n]$, and $A(q)$ and $C(q)$ are the AR and MA polynomials. 

When a FO is present in the system output, it is well-modeled as the response to a sinusoidal exogenous input to an ARX system (even if the actual source is technically not exogenous to the power system). Define system input $u$ as
\begin{equation}
	u[k] =  A\cos\left(2\pi \frac{f}{f_s} k + \theta\right)I_{\epsilon, \eta}[k]
	\label{eqn:FO}
\end{equation}
where $f_s$ is the sampling rate, $A$, $f$ and $\theta$ are the amplitude, frequency (Hz), and phase, and indicator function $I$ defines samples $\epsilon$ and $\eta$ where $u$ starts and ends, respectively.
\begin{equation}
	I_{\epsilon,\eta}[k] = \begin{cases}
				1, \quad \epsilon \leq k \leq \eta \\
				0, \quad \text{else} \\
			\end{cases}
\end{equation} 
The system response to $u$ is defined as $s$
\begin{equation}
	s[k] = \frac{B(q)}{A(q)}u[k] = \tilde{u}[k] + r[k]
	\label{eqn:Xdefn} 
\end{equation}
where B(q) is the X polynomial, $\tilde{u}$ is the FO observed in the system output 
\begin{equation}
	\tilde{u}[k] =  \tilde{A}\cos\left(2\pi \frac{f}{f_s} k + \tilde{\theta}\right)I_{\epsilon, \eta}[k]
	\label{eqn:FOhat}
\end{equation}
and $r$ is transient due to the FO (dis)appearing. It follows that the total observed system output $y$ is an ARMAX process
\begin{equation}
	y[k] = \frac{C(q)}{A(q)}e[k] + \frac{B(q)}{A(q)}u[k] = x[k] + s[k]
	\label{eqn:ARMAXdefn} 
\end{equation}
Note that $u$ is easily extended to a sum of multiple cosines.

\subsection{Changepoint Detection}
CPD is essentially a model selection problem in which a dataset is broken into segments defined by the CPs, each with their own parameters based on an assumed model, e.g., Gaussian processes with the same variance but different means \cite{truong}. Here it is assumed that both the number of CPs and their locations are unknown. CPD methods involve three pieces: the cost function, the search method, and the constraint on the number of CPs \cite{truong}. Mathematically, this is represented as a penalized cost function
\begin{equation}
	\min_{\tau}J(y,\tau) + p(\tau)
	\label{eqn:penCost}
\end{equation} 
where $\tau = \begin{bmatrix} \tau_1 & \ldots & \tau_m \end{bmatrix}$ is the vector of $m$ CPs. Referring to the aforementioned three choices, $J(y,\tau)$ is the cost function that is minimized by a search method, and $p(\tau)$ is the penalty function that constrains $m$.

The choice of cost function is depends on the underlying model assumption. For the problem of estimating the start and end of FOs, detection of mean-shifts was proposed in \cite{GM22} and used in \cite{GM23} as part of an iterative mode meter algorithm. The associated cost function is
\begin{equation}
	J(y,\tau) = \sum_{k=0}^{\tau_1-1}(y[k]-Y_0)^2 +\ldots+  \sum_{k=\tau_m}^{N-1}(y[k]-Y_m)^2
	\label{eqn:Cost}
\end{equation}
where the signal $y$ is broken into $m+1$ segments defined by the samples in $\tau$, and the $i^{th}$ segment has mean $Y_i$ \cite{truong,Kay}. 

There exist myriad search methods for solving the penalized cost function that may be broken into two categories: optimal and approximate \cite{truong}. In the latter category, the three most common are Binary Segmentation, Bottom Up, and Window Sliding, and each trade accuracy for speed \cite{truong}. The optimal approach, Pruned Exact Linear Time (PELT) \cite{killick2012optimal}, finds an exact solution to (\ref{eqn:penCost}) that begins with the idea that one could exhaustively iterate through every possible partitioning of $y$ (a slow procedure with computational cost $\mathcal{O}(N^2)$) and pick the one that minimizes the cost. PELT uses dynamic programming along with a pruning scheme that results in a drastic speed-up to $\mathcal{O}(N)$. PELT was used in \cite{GM22} and \cite{GM23} for the FO mode meter problem, and is used in the methods proposed here.

Finally, the penalty term must be chosen. While there are many choices in the literature, PELT requires that
\begin{equation}
	p(\tau) = \beta m
	\label{eqn:pen}
\end{equation}
where $\beta>0$ is the smoothing parameter \cite{truong, killick2012optimal, haynes2017}. The selection of $\beta$ is critical: too small and spurious changepoints are detected, too large and no changepoints are found. When the data follows a basic model, e.g., Gaussian, there are defined choices for $\beta$ \cite{truong}. However, as is the case with the data in the FO mode meter problem, $\beta$ must be tuned. The authors of \cite{haynes2017} propose the Changepoints for a Range of PenaltieS (CROPS) algorithm to iteratively run PELT over a shrinking range of $\beta$ values. While CROPS has the benefit of being agnostic to underlying models of $y$, it is computationally intensive as it calls PELT several times. 

The authors of \cite{truong} have made available the PELT, Binary Segmentation, Bottom Up, and Window Sliding search methods along with several cost functions in the Python Ruptures library \cite{ruptures}. The authors of \cite{killick2012optimal} have made their PELT algorithm available in both R \cite{r_pelt} and MATLAB \cite{matlab_pelt}. The reader is directed to the above reference for implementation details beyond those discussed below.

\subsection{Detecting When a Forced Oscillation Starts and Ends}
In \cite{GM22}, the problem of estimating the samples in $y$ where a FO starts and ends was derived as a mean-shift detection problem using the product-to-sum identity:
\begin{equation}
	2\cos(\theta_1)\cos(\theta_2) = \cos(\theta_1+\theta_2)+\cos(\theta_1-\theta_2)
\end{equation}
Define $y_{cos}= y\tilde{u}^*$ with $\tilde{u}^*$ being the output FO \textit{without} its indicator function such that it exists for all $N$ samples. Where the FO was not present in $y$, a zero-mean signal results:
\begin{equation}
	y_{cos}[k] = x[k]\tilde{A}\cos\left(2\pi \frac{f}{f_s} k + \tilde{\theta}\right)
	\label{eqn:ycos}
\end{equation}
However, where the FO is present, $y_{cos}$ becomes
\begin{equation}
\begin{split}
	y_{cos}[k] = & (x[k]+r[k])\tilde{A}\cos\left(2\pi \frac{f}{f_s} k + \tilde{\theta}\right)\\
				& + 0.5\tilde{A}^2\cos\left(2\pi \frac{2f}{f_s} k + \tilde{2\theta}\right)+ 0.5\tilde{A}^2 
\end{split}
\end{equation}
which has a mean of $0.5\tilde{A}^2$. Thus, the CPD problem is to find the samples at which $y_{cos}$ changes mean. 

This is achieved by first estimating the FO amplitude, frequency, and phase from the entirety of $y$ using the refined Discrete Fourier Transform method for $\hat{f}$ and then evaluating the Discrete-Time Fourier Transform at $\hat{f}$ to obtain $\hat{A}$ and $\hat{\theta}$ \cite{GM23}. These are used to construct an estimate of $\hat{y}_{cos}$ that is applied to MATLAB's \texttt{ischange} function with the \texttt{MaxNumChanges} option, which returns at most the user-specified number of CPs, $N_{maxCP}$. Also supplied by the user is a lower limit on segment size, $N_{minSL}$. Every time \texttt{ischange} finishes, if any segments are shorter than $N_{minSL}$, it is called again with a reduced $N_{maxCP}$. Thus, \texttt{ischange} is called at most $N_{maxCP}$ times until either no CPs are found or no segments are shorter than $N_{minSL}$. Once finished, the results are converted to values of $\epsilon$ and $\eta$ using a user-provided threshold $\alpha$ on the segment means. 

In \cite{GM23}, the method was implemented as part of a mode meter, and while estimation accuracy was excellent, the  algorithm was very slow since for every one call of the FO start/stop estimation algorithm, there are several calls of \texttt{ischange}. Compounding the issue is that when used with the \texttt{MaxNumChanges} option, the \texttt{ischange} function implements PELT with the CROPS algorithm to determine $\beta$, a process that calls MATLAB's private internal PELT method dozens, if not hundreds of times as it searches for an optimal solution.

\section{An Improved FO Changepoint Detector}
This paper proposes an improved method in which MATLAB's private internal PELT method is only called once, thus drastically improving speed, while reducing the number of user-specified parameters from 3 to 1. 

\subsection{Manual calculation of $\beta$}
Using the \texttt{ischange} function with the \texttt{Threshold}  option\footnote{\texttt{Y=ischange(ycos,`mean','Threshold',beta)}} instead of \texttt{MaxNumChanges} uses a user-supplied $\beta$ instead of using CROPS, so the internal PELT method is only called once. Note from \cite{killick2012optimal, haynes2017, matlab_pelt} that the solution to the CPD problem implies that the penalized cost function will be less than  the cost function applied to the entire dataset
\begin{equation}
\begin{split}
	\sum_{k=0}^{\tau_1-1}(y[k]-Y_0)^2+ & \ldots+\sum_{k=\tau_m}^{N-1}(y[k]-Y_1)^2 + \beta m \\
	& < \sum_{k=0}^{N-1}(y[k]-\bar{y})^2 = J(y,[\cdot])
\end{split}
\label{eqn:costineq}
\end{equation}
which implies the upper limit on $\beta$ is found with $m$=1
\begin{equation}
	\beta(\tau_1) <  J(y,[\cdot]) - \sum_{k=0}^{\tau_1-1}(y[k]-Y_0)^2 - \sum_{k=\tau_1}^{N-1}(y[k]-Y_1)^2
\label{eqn:betamax}
\end{equation}
so $\beta_{max}=\max(\beta(\tau_1))$ and $0<\beta<\beta_{max}$. In the FO problem, there are typically very few CPs, and so a wide range of $\beta$ provide the same accurate CP estimates.  Based on many simulation studies, the authors suggest that a $\beta$ in the ``middle'' of that range such as $0.5\beta_{max}$ or the mean of $\beta(\tau_1)$ work well.

\subsection{Converting changepoints to $\epsilon$ and $\eta$}
\begin{figure}[!t]
	\centering
	\includegraphics[width=3.5in]{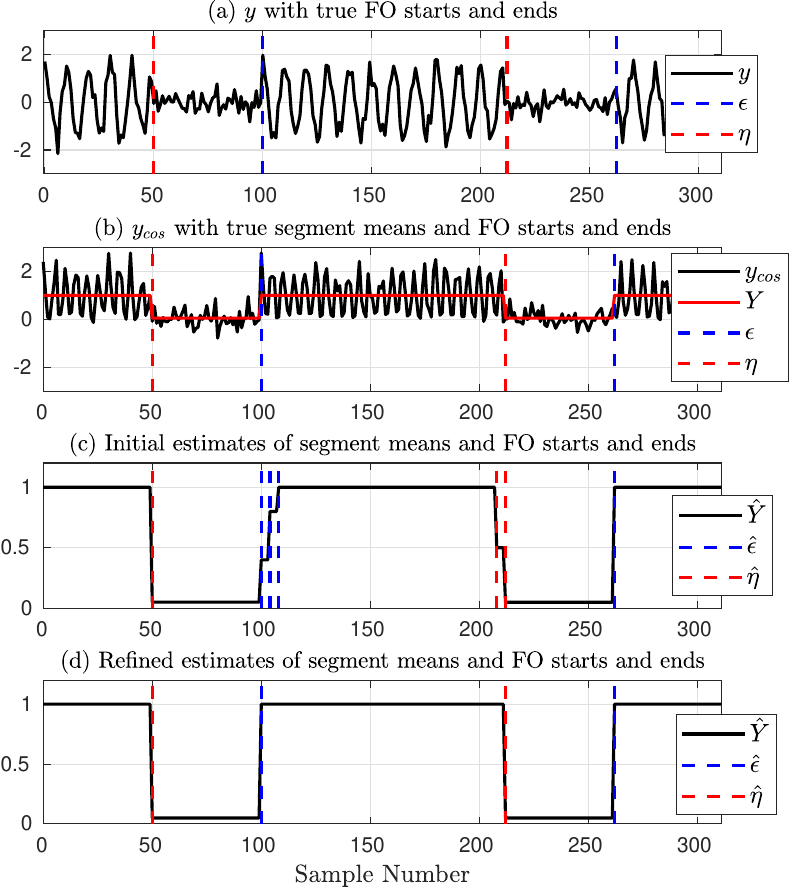}
	\caption{Improved usage of PELT to estimate FO start/stop samples.}
	\label{fig:NewWorkflow}	
	\vspace{-0.1in}
\end{figure}

While the method of \cite{GM22} used the estimated FO amplitude with a user-supplied scaling factor to obtain the FO start/end samples from the CPs, this paper proposes a method that relies on neither. First note that \texttt{ischange} returns $\hat{Y}$, an array the same size as $y_{cos}$ that is comprised of the estimated segment means, as seen in Fig. \ref{fig:NewWorkflow}. Let $dY$ be the backwards difference of $\hat{Y}$, and collect all indices  where $dY$ is positive as potential FO starting points in $\hat{\epsilon}=\begin{bmatrix} \hat{\epsilon}_{1} & \cdots & \hat{\epsilon}_{n_\epsilon}\end{bmatrix}$. Any indices where $dY$ is negative are potential FO endpoints collected as $\hat{\eta}=\begin{bmatrix} \hat{\eta}_{1} & \cdots & \hat{\eta}_{n_\eta}\end{bmatrix}$. 

Form array $\gamma$ as the sorted union of $\hat{\epsilon}$ and $\hat{\eta}$, e.g., Fig. \ref{fig:NewWorkflow}c, 
\begin{equation}
\gamma  = \begin{bmatrix} \hat{\eta}_{1} & \hat{\epsilon}_{1} & \hat{\epsilon}_{2} & \hat{\epsilon}_3 & \hat{\eta}_{2} & \hat{\eta}_{3} & \hat{\epsilon}_3  \end{bmatrix}
\end{equation}
where $\hat{\eta}_{1} < \hat{\epsilon}_{1} < \hat{\epsilon}_{2} < \hat{\epsilon}_3 < \hat{\eta}_{2} < \hat{\eta}_{3} < \hat{\epsilon}_3$. Ideally $\gamma$ does not have consecutive $\epsilon$ or $\eta$, resulting in a clear picture of when a FO is ``on'' and ``off.'' However, especially with low SNR, this may not be the case, so one can group consecutive $\epsilon$ by keeping the first in a sequence, and group consecutive $\eta$ by keeping the last in a group. Continuing the example and referring to Fig. \ref{fig:NewWorkflow}d, 
\begin{equation}
\begin{split}
\gamma  = &\begin{bmatrix} \hat{\eta}_{1} & \hat{\epsilon}_{1} & \xcancel{\hat{\epsilon}_{2}} & \xcancel{\hat{\epsilon}_3} & \hat{\eta}_{2} & \xcancel{\hat{\eta}_{3}} & \hat{\epsilon}_3  \end{bmatrix} \\
  \rightarrow &\begin{bmatrix} \hat{\eta}_{1} & \hat{\epsilon}_{1} &  \hat{\eta}_{2} &  \hat{\epsilon}_3  \end{bmatrix} \\
\end{split}
\end{equation}

Finally, $\gamma$ must be checked for situations where either a FO is already present at $k=0$ or when the FO is still going at $k=N-1$, as is the case in Fig. \ref{fig:NewWorkflow}. If the  proposed algorithm will return an $\eta$ as the first element of $\gamma$, a 0 must be appended to the start as an $\epsilon$. If the last element of $\gamma$ is an $\epsilon$, N-1 must be appended to the end as an $\eta$. For example,
\begin{equation}
\begin{split}
\gamma  = &\begin{bmatrix} \hat{\eta}_{1} & \hat{\epsilon}_{1} & \hat{\eta}_{2} &  \hat{\epsilon}_{2}  \end{bmatrix} \\
  \rightarrow  &\begin{bmatrix} 0 & \hat{\eta}_{1} & \hat{\epsilon}_{1} & \hat{\eta}_{2} &  \hat{\epsilon}_{2} & N\text{-1} \end{bmatrix} \\
\end{split}
\end{equation}
Once complete, $\gamma$ contain a sequence pairs ($\tilde{\epsilon}_i$, $\tilde{\eta}_i$) that define where the FO is `on', e.g., 
\begin{equation}
\begin{split}
\gamma  =  &\begin{bmatrix} 0 & \hat{\eta}_{1} & \hat{\epsilon}_{1} & \hat{\eta}_{2} &  \hat{\epsilon}_{2} & N\text{-1} \end{bmatrix} \\
  \rightarrow  &\begin{bmatrix} (\tilde{\epsilon}_1, \tilde{\eta}_{1}) & (\tilde{\epsilon}_2, \tilde{\eta}_{2}) &  (\tilde{\epsilon}_3, \tilde{\eta}_{3}) \end{bmatrix} \\
\end{split}
\label{eqn:finalCPs}
\end{equation}
where the tilde notation indicates a refined estimate.

\subsection{Minimum FO segment lengths}
Once the CP estimates have been refined in (\ref{eqn:finalCPs}), the user may wish throw away any FO ``on'' segments that are shorter than a particular threshold. The motivation here being that FOs below a particular SNR do not affect mode meters. Additionally, this could remove spurious estimated FO segments that were a result of noise.

Mode meters are sensitive to how much energy a FO has compared to the ARMA process at the FO frequency \cite{JimDetect}. The concept of a local SNR is used, 
\begin{equation}
	\text{SNR} = 10\log_{10}\left[ \frac{\frac{N_{FO}}{N}\frac{A^2}{2}}{\Phi_x(f)} \right]
	\label{eqn:localSNR}
\end{equation}
where $N_{FO}$ is the duration of the FO in samples, $N$ is the duration of the total signal $y$, $A$ is the FO amplitude, and $\Phi_x(f)$ is the power spectral density of the ARMA process $x$ at FO frequency $f$. Note that a short-duration, high-amplitude FO could have the same local SNR as a long-duration, low-amplitude FO at the same frequency. 

First determine the smallest SNR that biases an ARMA mode meter for a range of potential FO frequencies, $SNR_{min}$. Then, using domain expertise, determine the largest FO amplitude one might reasonably expect, $A_{max}$. Finally, use (\ref{eqn:localSNR}) to calculate a corresponding minimum FO segment length as
\begin{equation}
	N_{minSL} = \frac{2N10^{\frac{SNR_{min}}{10}}}{A^2_{max}}\Phi_x(f)
\end{equation}

\subsection{What if no CPs were detected?}
In the overall mode meter workflow, the FO time localization method is only called if the FO detection function returns a positive. Therefore, if no CPs are returned by the proposed algorithm, it could either mean the FO is present throughout the entirety of $y$, or there was a false positive returned from the detection algorithm. A simple way to test this is to create $\hat{x} = y-\tilde{u}^*$, where $\tilde{u}^*$ was the estimated FO used in (\ref{eqn:ycos}) to create $y_{cos}$. If FO $\tilde{u}$ is actually present in $y$, then subtracting $\tilde{u}^*$ will mostly cancel it out, resulting in an $\hat{x}$ that is a good approximation of the underlying ARMA process free from any FO. If however, there is not an FO present in $y$, creating $\hat{x}$ will actually introduce one in the form of $-\tilde{u}^*$. 

Apply the FO detection algorithm to $\hat{x}$. If nothing is detected, assume the FO is present throughout all of $y$ and set $\hat{\epsilon}=0$ and $\hat{\eta}=N$-1. Otherwise, assume that there is not a FO in $y$ and the original detection was a false positive.

\section{Simulation Study}
\label{sec:Sims}
To assess the proposed approach, a low-order ARMAX approximation of a single preprocessed output of the minniWECC model \cite{miniwecc} was used to generate bus frequency deviation data in units of mHz at 3 samples per second. The input GWN had variance of 0.16 to produce ambient data in same range as is observed in the real world. Each experiment simulated 25 minutes of data with a FO present for 10 minutes from samples 1535 to 3334. The FO amplitudes varied with SNR that ranged from -15 dB to 10 dB, and the FO frequency 0.370 Hz, chosen to be very close to the major 0.372 Hz, 4.67 \% damping North-South mode. For each of the cases studied below, 300 Monte Carlo trials were performed and means and standard deviations are reported.
\begin{figure}[!t]
	\centering
	\includegraphics[width=3.4in]{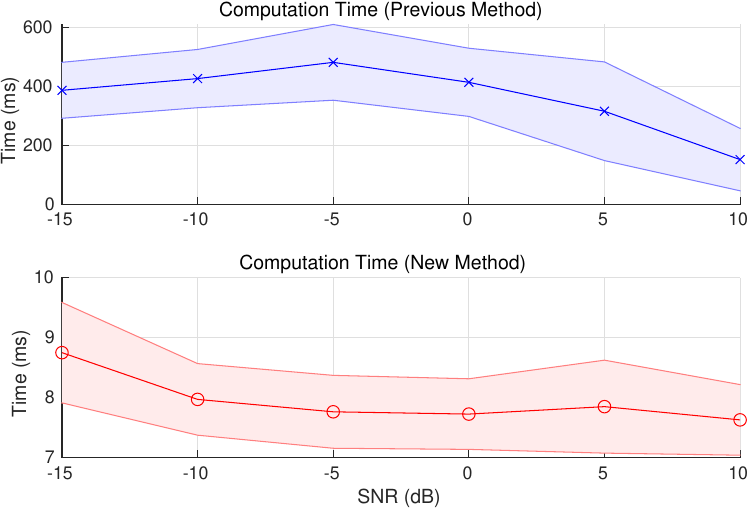}
	\caption{Means and regions of $\pm1$ std. dev of time to estimate $\epsilon$ and $\eta$.}
	\label{fig:timeCompare}	
	\vspace{-0.1in}
\end{figure}

For comparison purposes, the previous method of using PELT to estimate FO start and end points was implemented with the same settings as \cite{GM22}, $N_{maxCP}=10$, $N_{minSL}=2$ minutes, and $\alpha=0.7$. The proposed method was implemented using $\beta=\text{mean}(\beta(\tau_1))$. An 
ARMA mode meter bias study was conducted resulting in an $SNR_{min}$ of -15 dB, and the upper limit on FO amplitude, $A_{max}$, was chosen to be 10 mHz, slightly higher than the largest FO observed at Union College in Schenectady, NY. Together, this led to a minimum FO segment length $N_{minSL}$ of 36 samples, or 12 seconds.

The simulated data was subjected to the iterative mode meter work flow of \cite{GM23}. When detected, the FO amplitude, frequency, and phase are estimated to build $\hat{y}_{cos}$, which is sent to both the previous and proposed FO time localization algorithms. The outputs of each are used to get refined estimates of the FO amplitude, frequency, and phase that are passed on an ARMAX mode meter.

\begin{figure}[!t]
	\centering
	\includegraphics[width=3.4in]{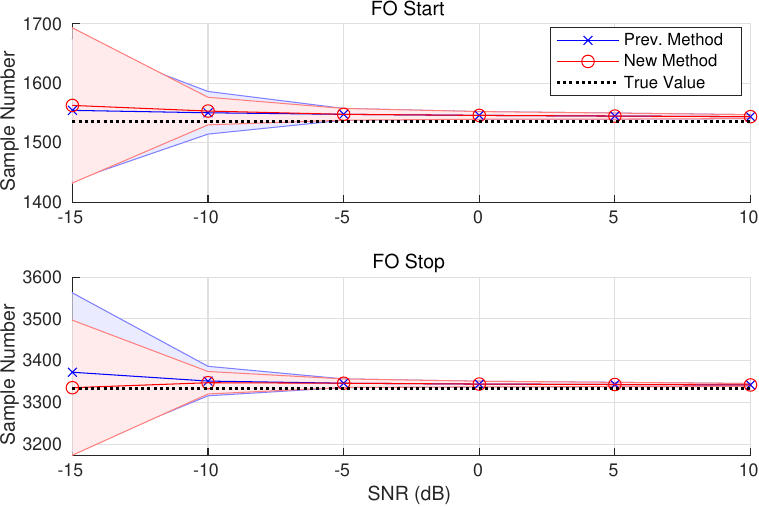}
	\caption{Means and regions of $\pm1$ std. dev of estimates of $\epsilon$ and $\eta$.}
	\label{fig:EpsEtaEstimates}	
	\vspace{-0.1in}
\end{figure}

As seen in Fig. \ref{fig:timeCompare}, the new algorithm enjoys a massive decrease in computational time. Across all SNR cases, the previous method had an average computational time of 363 ms while the proposed method was 8 ms. This 98\% reduction in computation time brings the FO start/stop estimation process down to the same order of magnitude as both the FO detection and FO amplitude, frequency, and phase estimation algorithms. As seen in Fig. \ref{fig:EpsEtaEstimates} despite being much faster, the proposed method retains the ability to accurately estimate $\epsilon$ and $\eta$, and even enjoys a slight reduction in variance at the lower SNRs.

\begin{figure}[!t]
	\centering
	\includegraphics[width=3.4in]{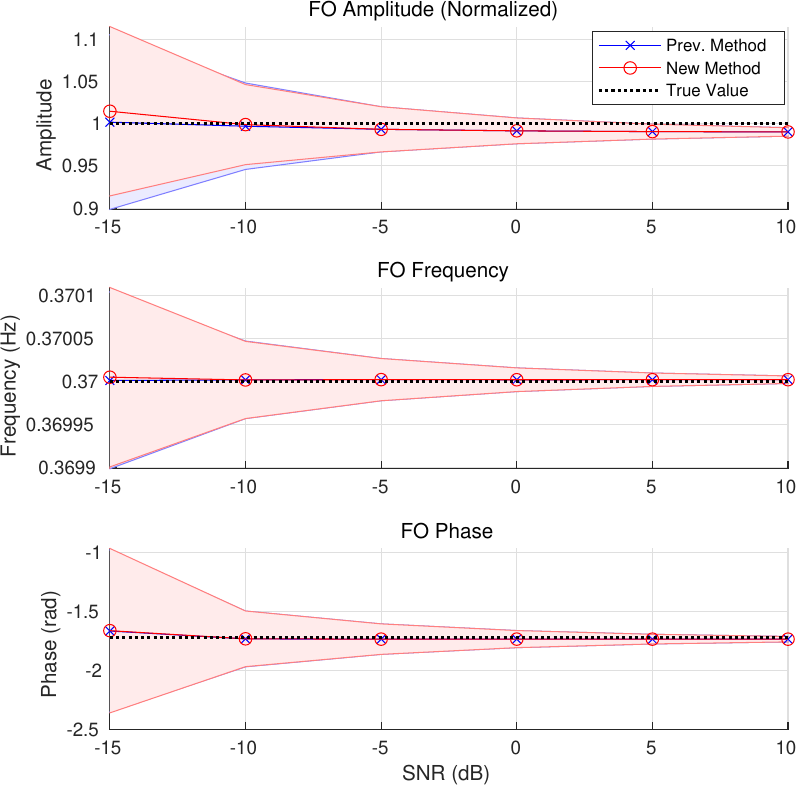}
	\caption{Means and regions of $\pm1$ std. dev of estimates of $A$, $f$, and $\theta$.}
	\label{fig:AFPhiEstimates}	
	\vspace{-0.2in}
\end{figure}

\begin{figure}[!t]
	\centering
	\includegraphics[width=3.4in]{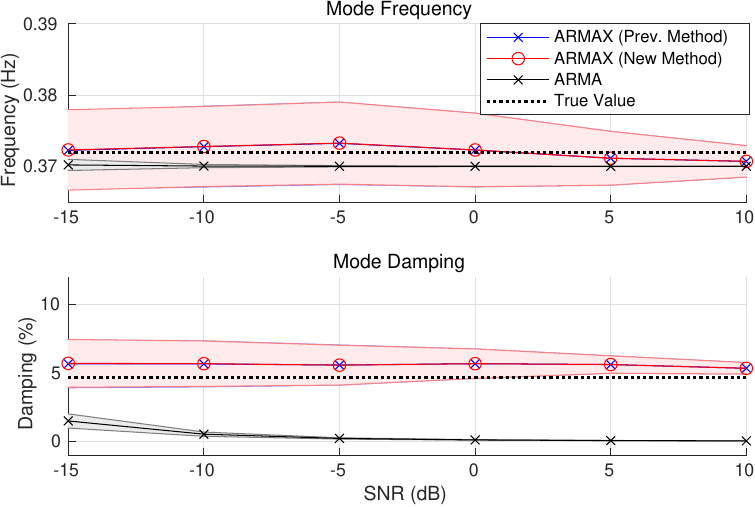}
	\caption{Means and regions of $\pm1$ std. dev of mode estimates.}
	\label{fig:ModeEstimates}	
	\vspace{-0.2in}
\end{figure}

Moving on to results that are not a direct output of the proposed algorithm, but rather depend upon them, Fig. \ref{fig:AFPhiEstimates} shows the refined estimates of FO amplitude, frequency, and phase obtained from analyzing only the portion of $y$ where the FO was estimated to be. Fig. \ref{fig:ModeEstimates} shows the results of the mode meter that used the estimated FOs as exogenous inputs (the highly biased ARMA mode meter results are included for comparison). The accuracy achieved by both methods are again nearly identical.

Several other scenarios were examined, e.g, including FOs of varying duration, FOs that either started before $k=0$, or FOs ended after $k=N$-1, and the results were the same - the proposed method was dramatically faster while maintaining estimation accuracy. Studies were performed using the Python Ruptures library to observe the effects of using Binary Segmentation, Bottom Up, or the Window Sliding methods instead of the PELT. Each of these demonstrated the unfortunate trade-off between speed and accuracy described in \cite{truong}. 

The presence of ringdown transients in the middle of the FO was considered. Small ringdowns relative to FO magnitude had no effect on either method, while large ringdowns caused the previous method to estimate the FO as present throughout the dataset, leading to biased mode estimates. The proposed method correctly estimated the FO start and end samples, but broke it into two segments: one that ends at the onset of the ringdown, and another that begins a few samples later. The mode estimates remained unbiased.

\section{Conclusions}
\label{sec:Conclusions}
This paper proposes a new method of estimating the start and end samples of FOs using CPD. Compared to the previous method, it enjoys a two-order-of-magnitude reduction in computation time without sacrificing accuracy. Future directions include investigating complex FO scenarios such as multiple or nonstationary FOs, and tests with real PMU data.

\bibliographystyle{IEEEtran}
\bibliography{IEEEabrv,PES_GM_2026}

\end{document}